\documentclass[conference]{IEEEtran}

\usepackage{cite}
\usepackage{amsmath,amssymb,amsfonts}
\usepackage{graphicx}
\graphicspath{{figures/}}
\usepackage{booktabs}
\usepackage{multirow}
\usepackage{url}
\usepackage[hypertexnames=false]{hyperref}
\usepackage{xcolor}
\usepackage{array}
\usepackage{algorithm}
\usepackage{algorithmic}
\usepackage{tikz}
\usetikzlibrary{shapes.geometric, arrows.meta, positioning, calc, backgrounds, fit}
\usepackage{listings}

\usepackage{stfloats}
\lstdefinestyle{agenticv2x}{
    basicstyle=\ttfamily\footnotesize,
    breaklines=true,
    breakatwhitespace=true,
    columns=fullflexible,
    keepspaces=true,
    showstringspaces=false,
    frame=single,
    framerule=0.25pt,
    rulecolor=\color{black!55},
    xleftmargin=0.6em,
    xrightmargin=0.6em,
    framexleftmargin=0.5em,
    framexrightmargin=0.5em,
    aboveskip=0.7em,
    belowskip=0.9em,
    captionpos=b
}

\title{Agentic-V2X: Small Language Model Agents for Deadline-Aware V2X Scheduling in 5G/6G Networks}

\author{
    \IEEEauthorblockN{Gerasimos Papanikolaou-Ntais\IEEEauthorrefmark{1}, Alexandros Kaloxylos\IEEEauthorrefmark{2}, and Athanasios Kanavos\IEEEauthorrefmark{2}}
    
    \IEEEauthorblockA{
        \IEEEauthorrefmark{1}Department of Informatics, University of Piraeus, 185 34 Piraeus, Greece \\
        \href{mailto:mpsp2540@unipi.gr}{mpsp2540@unipi.gr}
    }
    
    \IEEEauthorblockA{
        \IEEEauthorrefmark{2}Department of Informatics and Telecommunications, University of Peloponnese, 221 31 Tripoli, Greece \\
        \{\href{mailto:kaloxyl@uop.gr}{kaloxyl}, \href{mailto:kanavosa@uop.gr}{kanavosa}\}@uop.gr
    }
}

\begin{document}

\maketitle

\begin{abstract}
Large Language Models (LLMs) are increasingly proposed as control interfaces for next-generation networks. However, their latency, occasional hallucinations, and lack of control guarantees make them unsuitable for direct near-real-time packet schedulers. The problem is intensified in the highly dynamic environment of V2X communications. This paper studies a more practical architecture in which a small, locally deployed language model acts as a periodic non-real-time rApp-inspired policy creator, while a lightweight xApp-like controller executes validated policies at much faster intervals suitable for scheduling. The proposed framework targets deadline-aware 5G NR V2X scheduling with heterogeneous services, including teleoperated driving, cooperative awareness, HD map sharing, and sensor sharing. Given a scenario summary, service objective, and telemetry metrics, the LLM generates a structured scheduler policy containing service priorities, weight bounds, and safety constraints. A validator checks and repairs the policy before the xApp-like controller enforces it through scheduler-weight adaptation in ns-3/ns3-ai. The evaluation compares proportional fair scheduling, static expert policies, a heuristic xApp, static LLM-generated policies, and LLM-rApp policies with xApp-level enforcement over 126 completed runs. Metrics include deadline-constrained packet reception ratio, tail latency, deadline violations, throughput, fairness, policy validity, and safety interventions. Results show that the adaptive LLM-rApp/xApp design generates valid and executable policies throughout the campaign and remains competitive in several operating points, including improved mean critical reliability over PF at the highest density and favourable medium-density latency/throughput trade-offs. However, paired statistical analysis shows that the adaptive method is not the best aggregate critical-reliability method and remains below the strongest static policies overall. These results support Agentic-V2X as a safe and executable small-LLM-assisted policy-generation architecture rather than a universally dominant scheduler.
\end{abstract}

\begin{IEEEkeywords}
5G NR, V2X, O-RAN, rApp, xApp, small language models, ns-3, ns3-ai, scheduling, QoS, deadline-aware communication, network configuration.
\end{IEEEkeywords}

\section{Introduction}

Fifth-generation New Radio (5G NR) and emerging sixth-generation (6G) networks are evolving toward increasingly autonomous and AI-native architectures. The Open RAN (O-RAN) architecture has made this evolution explicit by separating control logic across two distinct timescales: non-real-time applications (rApps) hosted on the Non-RT RIC that reason over policies and objectives over seconds to minutes, and near-real-time applications (xApps) hosted on the Near-RT RIC that act on the radio over tens to hundreds of milliseconds~\cite{polese2023understanding}. This separation is attractive because it places slow, deliberative reasoning and fast, deterministic actuation in distinct functional roles.

Vehicle-to-everything (V2X) communications is one of the most demanding environments in next generation networks. A single cell must simultaneously serve several use cases, such as teleoperated driving (ToD)  with strict low-latency requirements, cooperative awareness messages, HD map distribution, and sensor sharing, each with highly different reliability, latency, and throughput needs~\cite{le2021v2x_rra_survey,gyawali2021cv2x_survey}. Configuring a scheduler to satisfy the strict latency needs of critical services while preserving background services' throughput is a well researched problem and has seen several works applying AI methods, most notably Reinforcement Learning, in order to optimize and balance this tradeoff.

Large language models (LLMs) have recently attracted attention for network management and orchestration, because they can interpret natural-language objectives, reason over heterogeneous service descriptions, and create structured configurations~\cite{telecom_llm_survey2024,bimo2025intent_ran_llm}. However, they are not suitable for direct integration into the near-real-time scheduling loop. Their inference latency is high and variable, and most importantly, their outputs are non-deterministic. They cannot offer guarantees of validity, safety, or bounded behavior. Using an LLM as the packet scheduler would couple the most safety-critical control loop in the RAN to the least predictable component in the architecture.

The key idea of this paper is therefore not to use the LLM as the scheduler. Instead, the LLM operates at the non-real-time policy layer, in a manner inspired by an rApp of the O-RAN architecture. It interprets a V2X service objective together with a periodic telemetry summary and produces a structured, actionable policy. A faster, deterministic xApp-like controller then executes this policy by adapting scheduler weights within predefined bounds, every 100~ms, while the LLM is consulted only on a slow timescale (every 10~s). All evaluation is performed in a simulation environment built on ns-3, 5G-LENA, SUMO, and ns3-ai. Because this setup is not a real deployed O-RAN rApp or xApp, we describe the components as rApp-inspired and xApp-like throughout this paper.

The central argument is that this division of labour is a more realistic and deployable use of LLMs in 5G/6G networks than direct online scheduling. LLMs are well suited to reason over service objectives, priorities, and policy trade-offs, while deterministic controllers are better suited to fast, repeated scheduling decisions with hard timing and safety requirements. The question we study is not whether an LLM can beat a hand-tuned controller on every metric, but whether a small, locally deployable LLM can generate valid, safe, executable, and competitive scheduling policies, and whether its integration is competitive to deployed and highly researched baselines.

\subsection*{Contributions}
This paper makes the following contributions.
\begin{enumerate}
    \item We propose \textbf{Agentic-V2X}, a simulation-based framework for small-LLM-assisted scheduler configuration in 5G NR V2X networks, in which a small local LLM acts as an rApp-inspired policy agent.
    \item We design a structured, rApp-inspired YAML policy interface together with validation, repair, and fallback mechanisms that prevent malformed or unsafe LLM outputs from directly affecting the scheduler.
    \item We implement a deterministic xApp-like executor that translates validated service-level policies into per-UE uplink/downlink scheduler weights through ns3-ai, operating at a 100~ms control period while the LLM is updated and generates the new (or keeps the existing) policy only every 10~s.
    \item We evaluate the framework across multiple V2X densities and seeds against proportional fair (PF), expert static baselines, a heuristic xApp, and static LLM-generated policies, analysing deadline-constrained PRR, reliability, deadline violations, background throughput, fairness, and policy validity/safety as an explicit trade-off rather than a single-winner comparison.
\end{enumerate}

We do not claim that the LLM universally outperforms deterministic baselines, as supported by the results, or that the system constitutes a certified or deployed O-RAN component. The contribution is the specific V2X-focused architecture and its reproducible evaluation.

\definecolor{colorNonRT}{RGB}{235, 243, 255}
\definecolor{colorNearRT}{RGB}{255, 244, 235}
\definecolor{colorSim}{RGB}{240, 249, 240}
\definecolor{colorLine}{RGB}{50, 50, 50}

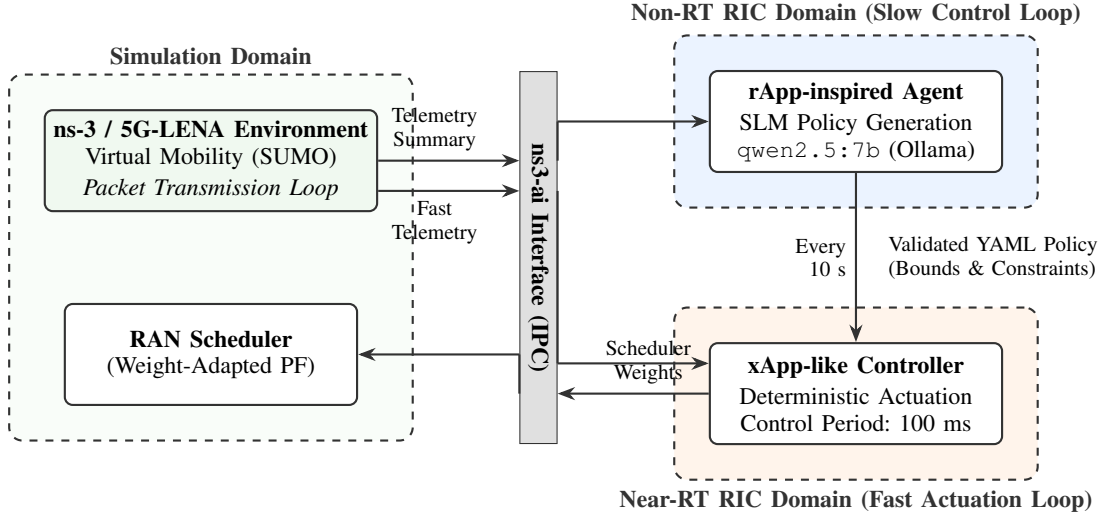
\begin{figure*}[!t]
\centering
\begin{tikzpicture}[
    node distance = 1.2cm and 2.2cm,
    box/.style = {rectangle, draw=colorLine, thick, fill=white, align=center, rounded corners=3pt, minimum height=3.8em, minimum width=11em, font=\small},
    container/.style = {rectangle, draw=colorLine, dashed, thick, rounded corners=5pt, inner sep=1.5em},
    arrow/.style = {->, >=Stealth, thick, draw=colorLine, text=black, font=\footnotesize}
]

    \node [box, fill=colorSim] (sim) {\textbf{ns-3 / 5G-LENA Environment}\\Virtual Mobility (SUMO)\\[0.2em]\textit{Packet Transmission Loop}};
    \node [box, fill=white, below=1.2cm of sim] (sched) {\textbf{RAN Scheduler}\\(Weight-Adapted PF)};
    
    \node [rectangle, draw=colorLine, fill=colorLine!15, minimum height=14em, minimum width=1.4em, right=2.0cm of $(sim.east)!0.5!(sched.east)$] (ipc) {};
    \node [rotate=270, font=\small\bfseries] at (ipc.center) {ns3-ai Interface (IPC)};

    \node [box, right=2.0cm of ipc, yshift=1.8cm] (rapp) {\textbf{rApp-inspired Agent}\\[0.2em]SLM Policy Generation\\\texttt{qwen2.5:7b} (Ollama)};
    \node [box, right=2.0cm of ipc, yshift=-1.8cm] (xapp) {\textbf{xApp-like Controller}\\[0.2em]Deterministic Actuation\\Control Period: 100~ms};

    \draw [arrow] (sim.east) -- node[above, align=center, pos=0.4] {Telemetry\\Summary} (ipc.west |- sim.east);
    \draw [arrow] (ipc.east |- rapp.west) -- (rapp.west);
    \draw [colorLine, thick] (ipc.east |- sim.east) -- (ipc.east |- rapp.west); 
    
    \draw [arrow] (rapp.south) -- node[right=0.3cm, align=left] {Validated YAML Policy\\(Bounds \& Constraints)} node[left, align=right]{Every\\10~s} (xapp.north);

    \draw [arrow] ($(sim.east)+(0,-0.4)$) -- node[below, align=center, pos=0.4] {Fast\\Telemetry} ($(ipc.west |- sim.east)+(0,-0.4)$);
    \draw [arrow] ($(ipc.east |- xapp.west)+(0,0.4)$) -- ($(xapp.west)+(0,0.4)$);
    \draw [colorLine, thick] ($(ipc.east |- sim.east)+(0,-0.4)$) -- ($(ipc.east |- xapp.west)+(0,0.4)$); 

    \draw [arrow] (xapp.west) -- node[above, align=center, pos=0.4] {Scheduler\\Weights} (ipc.east |- xapp.west);
    \draw [arrow] (ipc.west |- sched.east) -- (sched.east);
    \draw [colorLine, thick] (ipc.west |- xapp.west) -- (ipc.west |- sched.east); 

    \begin{scope}[on background layer]
        \node [container, fill=colorNonRT, fit=(rapp), inner sep=1.3em, label={[font=\small\bfseries, text=colorLine]above:Non-RT RIC Domain (Slow Control Loop)}] (boxNonRT) {};
        \node [container, fill=colorNearRT, fit=(xapp), inner sep=1.3em, label={[font=\small\bfseries, text=colorLine]below:Near-RT RIC Domain (Fast Actuation Loop)}] (boxNearRT) {};
        \node [container, fill=colorSim!40, fit=(sim) (sched), inner sep=1.3em, label={[font=\small\bfseries, text=colorLine]above:Simulation Domain}] (boxEnv) {};
    \end{scope}

\end{tikzpicture}
\caption{Proposed Agentic-V2X architecture. The system establishes strict timescale separation, isolating the stochastic small language model inside a slow non-real-time policy-generation loop while a deterministic xApp controller handles near-real-time scheduler adaptation via the ns3-ai module.}
\label{fig:architecture}
\end{figure*}

\section{Related Work}

\subsection{V2X Scheduling and Radio Resource Management}
Radio resource management for V2X has been studied extensively under both cellular (Uu) and sidelink (PC5) interfaces, with a recurring emphasis on meeting latency and reliability requirements for safety-critical traffic while sharing resources with best-effort flows~\cite{le2021v2x_rra_survey,gyawali2021cv2x_survey}. Classical schedulers such as proportional fair (PF) operate by balancing throughput and fairness but are not deadline-aware, and QoS-aware extensions typically introduce per-flow weighting, priority queues, or delay-sensitive metrics~\cite{le2021v2x_rra_survey}. In the 5G NR setting, the introduction of flexible numerologies and configurable schedulers has increased the configuration space and motivated automated or policy-driven weight selection~\cite{gyawali2021cv2x_survey}. Our work sits at this configuration layer: rather than proposing a new scheduling discipline or algorithm, we study how service-level priorities (or policies) can be turned into per-UE UL/DL weights for an otherwise standard PF-based scheduler with weight adaptation capabilities.

\subsection{AI/RL for V2X Resource Allocation}
Machine learning, and reinforcement learning (RL) in particular, has been applied to V2X resource allocation, power control, and mode selection, often demonstrating gains over static heuristics in simulation~\cite{ye2018drl_v2v_icc,ye2019drl_v2v,liang2019spectrum_marl}. These approaches, however, generally require substantial training, careful reward shaping, and retraining when the service mix or environment changes, and they can be difficult to constrain for safety~\cite{ye2019drl_v2v,liang2019spectrum_marl}. The framework proposed here is complementary: the deterministic xApp-like controller occupies the fast control loop where learned online control is most risky, while the slow, language-model policy layer supplies objectives and bounds without requiring task-specific online training.

\subsection{LLMs and Agentic Methods for RAN/O-RAN Management}
A growing body of work explores the integration of LLMs in telecom and RAN management, from standards question answering and retrieval-augmented generation over specifications~\cite{telecom_llm_survey2024,bariah2023telecom_language}, intent translation for intent-based networking~\cite{dzeparoska2023llm_policy,bimo2025intent_ran_llm}, configuration generation, troubleshooting, to broader O-RAN orchestration~\cite{llm_hric2025}. Several efforts frame the LLM as an agent that plans, calls tools, or coordinates network functions~\cite{xu2024llm_agents_6g,wang2024llm_agents_survey}. Most of this literature targets management-plane reasoning and human-facing automation rather than the scheduling control loop. Where LLMs have been considered closer to control, the high and variable inference latency and the lack of output guarantees are widely recognised obstacles~\cite{xu2024llm_agents_6g,telecom_llm_survey2024}. We adopt the same diagnosis and respond architecturally: the LLM is confined to slow policy generation, and a separate deterministic component performs all fast actuation.

\subsection{rApps, xApps, and Hierarchical Control}
The O-RAN architecture formalises a separation between non-real-time control on the Non-RT RIC (rApps, operating over seconds to minutes) and near-real-time control on the Near-RT RIC (xApps, operating over tens to hundreds of milliseconds), connected through the A1 and E2 interfaces~\cite{polese2023understanding}. This hierarchical structure is a natural fit for combining deliberative reasoning with fast actuation. Our design integrates this structure conceptually: the LLM plays an rApp-inspired role as a slow policy creator, and the deterministic controller plays an xApp-like role as a fast policy executor. We emphasize that our implementation is simulation-based and does not implement A1/E2 interfaces, a real Non-RT or Near-RT RIC, or a conformant rApp/xApp; the terminology is used to describe the timescale separation, not a deployment.

\subsection{Positioning of This Work}
This paper evaluates whether a small, locally deployable LLM can act as an rApp-inspired policy creator whose structured outputs are validated, repaired if necessary, and enforced by a deterministic xApp-like controller. Table~\ref{tab:related_work_positioning} contrasts this direction with adjacent lines of work.

\begin{table}[!tb]
\centering
\caption{Positioning of the proposed work against related LLM-for-networking directions.}
\label{tab:related_work_positioning}
\begin{tabular}{p{0.27\linewidth} p{0.27\linewidth} p{0.32\linewidth}}
\toprule
\textbf{Direction} & \textbf{Typical Focus} & \textbf{Difference of This Work} \\
\midrule
Telecom RAG & Standards Q\&A & Executable V2X scheduler policies evaluated in ns-3 \\
Intent-based RAN & Intent translation & rApp-inspired LLM policy creation with xApp-like enforcement \\
LLM agents & General network automation & Small local LLM with constrained policy validation \\
RL scheduling & Learned online control & Fast deterministic controller; LLM only creates policies \\
O-RAN xApps & Near-real-time control & LLM supplies validated policy bounds/objectives \\
\bottomrule
\end{tabular}
\end{table}

\section{System Model and Problem Formulation}

\subsection{5G NR V2X Scenario}
\begin{figure}[!tb]
    \centering
    \includegraphics[width=0.90\linewidth]{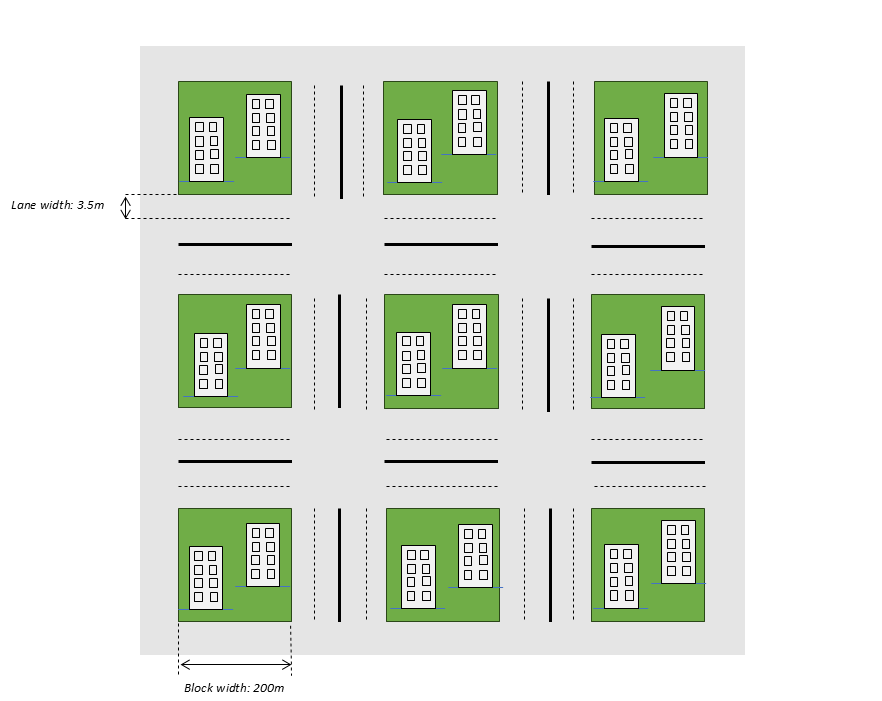}
    \caption{Manhattan-style Urban grid}
    \label{fig:manhattan}
\end{figure}

We evaluate our architecture using a single 5G NR cell served by one, centrally placed gNB, with vehicular mobility generated by SUMO~\cite{lopez2018sumo} on a Manhattan-style grid, as shown in Figure \ref{fig:manhattan}. The radio layer is simulated using ns-3 with the 5G-LENA module~\cite{patriciello2019lena,koutlia2022lena_calibration}, and the external control processes are connected to the simulator through ns3-ai~\cite{yin2020ns3ai}. Three vehicle densities are studied: 20, 25, and 30 vehicles, and each density experiment is repeated over seven random seeds. The offered load per vehicle is fixed and compliant to the V2X 5GAA application definitions, so increasing vehicle density increases aggregate offered load and channel contention because more vehicles generate traffic according to the same fixed service models.

\subsection{V2X Service Classes and Deadlines}
Four V2X services are evaluated, with two critical and two background classes. Teleoperated driving (ToD) is a critical, strict low-latency service whose uplink carries a high-rate video/telemetry stream and whose downlink carries low-rate control. Cooperative awareness (referred to as Awareness/AIC) is a critical, low-rate, bidirectional service. HD map distribution (HDMap/HDM) is a background service dominated by a high-rate downlink. Sensor sharing (RTSA) is a background, moderate-rate bidirectional service. Table~\ref{tab:v2x_services} lists the per-service traffic rates, criticality, direction, and the deadline used to classify a packet as on-time. The deadline values are scenario design parameters used to compute deadline-constrained reception and are not experimental outcomes.

\begin{table}[!tb]
\centering
\caption{V2X service classes, traffic rates, criticality, and deadline relevance.}
\label{tab:v2x_services}
\begin{tabular}{l c c c c}
\toprule
\textbf{Service} & \textbf{Criticality} & \textbf{Dir.} & \textbf{Rate} & \textbf{Deadline} \\
\midrule
ToD (UL)      & Critical   & UL    & 16 Mbps     & 20 ms \\
ToD (DL)      & Critical   & DL    & 0.048 Mbps  & 20 ms \\
Awareness/AIC & Critical   & UL/DL & 0.032 Mbps  & 100 ms \\
HDMap (DL)    & Background & DL    & 16 Mbps     & 1000 ms \\
HDMap (UL)    & Background & UL    & 0.008 Mbps  & 1000 ms \\
Sensor/RTSA   & Background & UL/DL & 0.048 Mbps  & 500 ms \\
\bottomrule
\end{tabular}
\end{table}

The per-density service mixes (number of flows of each type) are summarised in Table~\ref{tab:experiment_matrix} and increase with density, so that higher densities stress both the critical and background classes more heavily.

\subsection{Scheduler-Weight Configuration}
The scheduler is PF-based with configurable per-service weights, which are mapped to per-UE UL/DL scheduler weights inside ns-3. The control problem is hierarchical and operates at two timescales. A slow rApp-inspired policy agent produces a policy $\pi_{\mathrm{rApp}}$ containing baseline weights, allowed weight ranges, service priorities, and safety constraints. A fast xApp-like executor produces $\pi_{\mathrm{xApp}}$, the sequence of concrete weight settings applied over time, by adapting weights within the validated bounds in response to observed telemetry. The scheduler then applies the resulting weights to dynamically handle resource allocation. Crucially, the LLM output is not a direct scheduler action, and the xApp action is always bounded by the validated policy constraints.

\subsection{Objective}
The design goal is to maximise critical-service reliability (ToD and Awareness deadline-constrained reception) and minimize deadline violations, while preserving background (HDMap/Sensor) throughput, avoiding resource starvation of any service, and avoid ever applying an unsafe or invalid policy. These goals are partly conflicting: prioritising the high-rate critical flows consumes resources that would otherwise serve background traffic. We do not combine them into a single scalar objective, instead each goal corresponds to one or more of the metrics reported in Section~\ref{sec:metrics} (critical DC-PRR, deadline-violation rate, background throughput, fairness as a starvation indicator, and safety interventions), on which the methods are compared individually, so that the resulting trade-offs are visible rather than collapsed into a single metric.

\section{Agentic-V2X Framework}

\subsection{Framework Overview}
Agentic-V2X separates slow policy reasoning from fast policy execution, as shown in Fig.~\ref{fig:architecture}. The pipeline is: operator objective and scenario summary $\rightarrow$ small local LLM (rApp-inspired agent) $\rightarrow$ structured YAML policy $\rightarrow$ validator (with repair and fallback) $\rightarrow$ deterministic xApp-like controller $\rightarrow$ ns3-ai $\rightarrow$ ns-3/5G-LENA scheduler $\rightarrow$ telemetry, optionally fed back to the next policy update. The LLM-based agent receives an operator objective and a compact scenario summary and produces a policy specifying baseline service weights, allowed ranges, adaptation hints, and safety constraints. This policy is validated before being handed to the controller. The controller uses no natural language: it observes telemetry such as latency, deadline-violation rate, throughput degradation, and congestion indicators, and adjusts scheduler weights within the bounds set by the validated policy. If the policy is invalid, unsafe, or missing required fields, the controller falls back to the last known safe policy or to a static expert baseline.

\subsection{Small Language Model Policy Agent}
We use the term \emph{agent} in the sense of an autonomous reasoning component that ingests a telemetry summary and service objective and emits an actionable, structured policy within a closed perceive--decide--act loop, rather than in the stronger sense of a tool-calling or multi-step planning agent. Richer forms of agency, such as tool invocation or iterative planning, are natural extensions that we do not pursue here, in part because the ns-3/5G-LENA control environment does not expose a tool-calling interface to the policy layer, so we leave them to future work. We use \texttt{qwen2.5:7b} as a representative small local model. It is not claimed to be optimal, but rather, it was selected as a practical trade-off between local deployability and reliable generation of structured YAML policies. In preliminary tests, very small models such as \texttt{qwen2.5:0.5b} produced unreliable or poorly structured outputs. In addition, the output latency of larger models, such as a 14-billion-parameter model, would be prohibitive for deployment. The agent is queried on a slow, non-real-time interval, specifically once before the run for the static configuration, and afterwards, every 10~s for the adaptive configuration, so that LLM inference is never inside the scheduling loop. Its input includes the V2X service descriptions and deadlines, the current scenario telemetry summary (density and recent performance), and the operator objective. After inference, the xApp receives its output as a machine-checkable policy.

\subsection{Structured YAML Policy Schema}
The agent emits a structured YAML policy whose fields are summarised in Table~\ref{tab:policy_schema}. The schema constrains the LLM to a bounded, validatable action space: baseline weights and explicit per-service min/max bounds, safety constraints (such as a minimum weight for critical services and a maximum tolerated degradation of background throughput), optional adaptation rules expressed as guarded hints for the controller, an update period, and a named fallback policy.

\begin{table}[!tb]
\centering
\caption{Structured rApp-inspired policy schema fields.}
\label{tab:policy_schema}
\begin{tabular}{p{0.30\linewidth} p{0.60\linewidth}}
\toprule
\textbf{Field} & \textbf{Meaning} \\
\midrule
\texttt{policy\_id} & Unique identifier for the generated policy \\
\texttt{baseline\_weights} & Per-service starting weights (ToD, Awareness, HDMap, Sensor) \\
\texttt{bounds} & Per-service [min, max] weight ranges the controller may use \\
\texttt{safety\_constraints} & Min critical weight, max background degradation, anti-starvation \\
\texttt{xapp\_rules} & Optional guarded adaptation hints for the controller \\
\texttt{update\_period} & Slow policy refresh cadence (10~s) \\
\texttt{fallback\_policy} & Named safe policy used if validation fails \\
\texttt{runtime\_safety} & Rollback and critical UL/DL balancing parameters enforced by the xApp \\
\bottomrule
\end{tabular}
\end{table}

\subsection{Policy Validation, Repair, Fallback, and Runtime Shielding}

Because LLM outputs are non-deterministic, no generated policy can be trusted directly. Each policy is first parsed against the YAML schema and checked for structural completeness, admissible service and metric names, valid action types, weight and bound ranges, consistency between baseline weights and bounds, and satisfaction of safety constraints such as minimum critical-service weight and non-zero background weights. Recoverable problems, such as clampable out-of-range values or missing optional fields, are repaired deterministically and logged. Unrecoverable schema violations or unsafe policies are rejected before they can affect the scheduler.

In addition to structural validation, the adaptive LLM-rApp/xApp path includes a runtime performance-aware shield. When a new validated LLM policy is accepted, the xApp stores the previous validated policy and records the current critical-flow performance as a baseline. During subsequent control windows, the xApp monitors critical violation rate, critical DC-PRR, and critical p95 latency. If critical-service performance degrades for three consecutive windows relative to the baseline, the xApp rolls back to the previous validated policy. If no previous validated policy is available, the controller falls back to the static expert policy. This mechanism protects against policies that are syntactically valid and within bounds but harmful under the current traffic and channel state.

The xApp also applies a directional critical-flow guard. If the gap between critical UL and DL DC-PRR exceeds 0.05, the controller temporarily boosts the scheduler weights of the weaker critical direction until the imbalance improves. This guard is intended to avoid over-protecting one critical direction while allowing the other to degrade, especially under high-density congestion.

Thus, the safety design has two levels. The validator prevents malformed or out-of-bounds YAML policies from reaching the scheduler, while the runtime shield protects against validated but performance-harmful policies. In the worst case, the system degrades to either the last validated safe policy or the static expert fallback.

\begin{algorithm}[!tb]
\caption{Validation, Runtime Shielding, and xApp Enforcement}
\label{alg:validation_xapp}
\begin{algorithmic}[1]
\REQUIRE LLM-generated policy $p$, telemetry $z_t$, previous safe policy $p_{\mathrm{prev}}$
\ENSURE scheduler weights $w_t$
\STATE Parse $p$ against the YAML policy schema
\IF{$p$ is incomplete or violates a recoverable rule}
    \STATE Repair $p$ deterministically; log repair
\ENDIF
\IF{$p$ is invalid or violates a structural safety constraint}
    \IF{$p_{\mathrm{prev}}$ exists}
        \STATE Load previous validated policy $p_{\mathrm{prev}}$
    \ELSE
        \STATE Load static expert fallback policy
    \ENDIF
\ELSE
    \STATE Store current validated policy as $p_{\mathrm{prev}}$
    \STATE Accept new validated policy $p$
    \STATE Record current critical-flow performance baseline
\ENDIF
\STATE Observe telemetry $z_t$
\IF{critical violation rate, critical DC-PRR, or critical p95 latency degrades for 3 consecutive windows}
    \IF{$p_{\mathrm{prev}}$ exists}
        \STATE Roll back to $p_{\mathrm{prev}}$
    \ELSE
        \STATE Load static expert fallback policy
    \ENDIF
\ENDIF
\IF{$|\mathrm{DC\text{-}PRR}_{\mathrm{crit,UL}} - \mathrm{DC\text{-}PRR}_{\mathrm{crit,DL}}| > 0.05$}
    \STATE Temporarily boost weights of the weaker critical direction
\ENDIF
\STATE Apply xApp adaptation within validated bounds and hysteresis limits
\STATE Enforce minimum critical weights and anti-starvation constraints
\STATE \textbf{return} scheduler weights $w_t$
\end{algorithmic}
\end{algorithm}

\subsection{Deterministic xApp-Like Execution}
\begin{figure}[!tb]
\centering
\setlength{\unitlength}{1mm}
\begin{picture}(86,34)

\put(3,25){\small rApp-inspired LLM: policy generation and validation every 10 s}
\put(5,20){\line(1,0){74}}
\put(5,18){\line(0,1){4}}
\put(30,18){\line(0,1){4}}
\put(55,18){\line(0,1){4}}
\put(79,18){\line(0,1){4}}
\put(3,14){\small 0 s}
\put(27,14){\small 10 s}
\put(52,14){\small 20 s}
\put(76,14){\small 30 s}

\put(3,8){\small xApp-like executor: deterministic weight adaptation every 100 ms}
\put(5,4){\line(1,0){74}}
\multiput(5,2)(2.5,0){30}{\line(0,1){4}}
\put(20,-1){\small 100 xApp steps per 10 s policy interval}

\end{picture}
\caption{Control timescales in Agentic-V2X. The LLM policy agent operates every 10~s; the deterministic xApp-like controller actuates every 100~ms, keeping LLM inference out of the scheduling loop.}
\label{fig:timescales}
\end{figure}
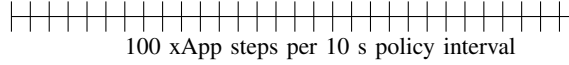

The xApp-like executor is a lightweight deterministic controller that runs every 100~ms. It enforces the validated policy by adjusting scheduler weights within the validated bounds in response to telemetry. For example, it may raise the ToD weight when ToD UL latency exceeds a threshold, restore HDMap weight when critical traffic is stable, and prevent any background service from being driven to zero resources. Hysteresis and rate limiting are applied to avoid oscillatory behaviour, and all actions are clamped to the policy bounds, so the controller cannot exceed the constraints supplied by the policy agent. The two control timescales, LLM policy updates every 10~s versus deterministic weight adaptation every 100~ms, are depicted in Fig.~\ref{fig:timescales}.

\section{Experimental Methodology}
\label{sec:methodology_label_placeholder}

\subsection{ns-3 / 5G NR Setup}
Experiments use ns-3 with the 5G-LENA NR module~\cite{patriciello2019lena,koutlia2022lena_calibration}, SUMO mobility on a Manhattan grid~\cite{lopez2018sumo}, and ns3-ai for external control~\cite{yin2020ns3ai}. The scheduler is PF-based with configurable per-UE UL/DL weights. The xApp-like control period is 100~ms; the rApp-inspired LLM update period is 10~s for the adaptive method. Table~\ref{tab:simulation_parameters} lists the main simulation parameters.

\begin{table}[!tb]
\centering
\caption{Simulation parameters.}
\label{tab:simulation_parameters}
\begin{tabular}{l c}
\toprule
\textbf{Parameter} & \textbf{Value} \\
\midrule
Simulator & ns-3 / 5G-LENA \\
External control & ns3-ai \\
Mobility & SUMO Manhattan grid \\
Densities & 20, 25, 30 vehicles \\
Seeds & 7 seeds per density \\
Scheduler & PF with configurable per-UE UL/DL weights \\
xApp control period & 100 ms \\
rApp update period & 10 s (adaptive method) \\
LLM serving & Ollama, \texttt{qwen2.5:7b} \\
Policy format & YAML (validated/repaired) \\
Primary metric & Critical DC-PRR \\
\bottomrule
\end{tabular}
\end{table}

\subsection{Traffic and Service Mix}
The four services of Table~\ref{tab:v2x_services} are instantiated according to the per-density mixes in Table~\ref{tab:experiment_matrix}. Per-flow rates are fixed across densities, so aggregate offered load and contention scale with the number of vehicles. The high-rate flows (ToD UL and HDMap DL, each 16~Mbps) dominate the resource demand and create the principal tension between critical reliability and background throughput that the scheduler-configuration policies must manage.

\subsection{Baselines}
We compare six scheduler-control methods, summarised in Table~\ref{tab:methods}. The first four are deterministic references while the last two use the LLM policy agent, differing in whether the policy is static or adaptively re-generated and executed.

For readability, the exact implementation identifiers are mapped to reader-friendly method names as follows: static expert (\texttt{static\_expert}), balanced expert (\texttt{balanced\_expert}), heuristic xApp (\texttt{heuristic\_xapp}), static LLM (\texttt{llm\_static}), and adaptive LLM-rApp/xApp (\texttt{llm\_rapp\_xapp}). We use the reader-friendly method names in the remainder of the paper.

\begin{table}[!tb]
\centering
\caption{Compared methods and their control behaviour.}
\label{tab:methods}
\begin{tabular}{p{0.27\linewidth} p{0.62\linewidth}}
\toprule
\textbf{Method} & \textbf{Description} \\
\midrule
PF & True PF baseline; all service weights equal to 1 with no prioritisation \\
static expert & Deterministic, manually selected static service weighting favouring critical traffic \\
balanced expert & Deterministic, manually selected static weighting balancing critical and background \\
heuristic xApp & Deterministic adaptive xApp-like controller, no LLM policy generation \\
static LLM & LLM generates one validated policy before the run, static weights throughout \\
LLM-adaptive & LLM generates/updates a validated policy every 10~s, deterministic xApp-like controller executes it every 100~ms \\
\bottomrule
\end{tabular}
\end{table}

\subsection{Experiment Matrix}
\label{sec:experiment_matrix}
The campaign is a full factorial of 3 densities $\times$ 7 seeds $\times$ 6 methods $=$ 126 runs, all of which completed successfully and are included in the analysis. Table~\ref{tab:experiment_matrix} lists the densities, per-density service mixes, seeds, and methods. The same seven mobility seeds are reused across all six methods within each density, so that comparisons are made on identical vehicle trajectories and traffic instantiations, and every method is evaluated over the same 21 runs (three densities $\times$ seven seeds).

\begin{table}[!tb]
\centering
\caption{Experiment matrix and per-density service mixes.}
\label{tab:experiment_matrix}
\begin{tabular}{l c c c c}
\toprule
\textbf{Density} & \textbf{HDM} & \textbf{AIC} & \textbf{ToD} & \textbf{RTSA} \\
\midrule
20 vehicles & 1 & 5 & 1 & 13 \\
25 vehicles & 2 & 6 & 2 & 15 \\
30 vehicles & 3 & 7 & 3 & 17 \\
\midrule
\multicolumn{5}{l}{Seeds: 1, 2, 3, 4, 5, 6, 7} \\
\multicolumn{5}{l}{Methods: PF, static expert, balanced expert,} \\
\multicolumn{5}{l}{\quad heuristic xApp, static LLM, adaptive LLM-rApp/xApp} \\
\multicolumn{5}{l}{Total runs: $3 \times 7 \times 6 = 126$} \\
\bottomrule
\end{tabular}
\end{table}

\subsection{Metrics}
\label{sec:metrics}
We report network-level and control-level metrics. Network metrics: critical deadline-constrained packet reception ratio (critical DC-PRR, combining ToD and Awareness), per-service DC-PRR (ToD downlink/uplink, Awareness downlink/uplink), deadline-violation rate, background (HDMap + Sensor/RTSA) throughput and degradation, 95th-percentile (p95) latency, and the Jain fairness index over per-service throughput. Control/policy metrics: policy validity, repair, and rejection counts, fallback invocations, number of xApp weight-change actions, number of safety interventions, LLM policy-generation latency, and xApp per-step execution overhead. The primary metric is critical DC-PRR; the others characterise the reliability--throughput--fairness trade-off and the practicality of the policy agent.


\section{Results}
This section reports results from the complete 126-run experimentation process (3 densities $\times$ 7 seeds $\times$ 6 methods). All figures are computed from the per-run telemetry described in Section~\ref{sec:metrics}; per-method aggregates are means over the seven seeds at each density. Because the same density/seed combinations are reused across all methods, the main comparisons are paired rather than treated as independent samples.

\subsection{Policy Validity and Safety}
\label{sec:results_validity}
\begin{table}[!tb]
\centering
\caption{Policy validity and safety outcomes (21 runs per LLM-based method; adaptive LLM-rApp/xApp issues 10 policy updates per run, 210 total).}
\label{tab:policy_validation}
\begin{tabular}{l c c c c}
\toprule
\textbf{Method} & \textbf{Valid/accepted} & \textbf{Repaired} & \textbf{Rejected} & \textbf{Fallback} \\
\midrule
static LLM & 21/21 (100\%) & 0 & 0 & 6/21 \\
LLM-adaptive & 210/210 (100\%) & 0 & 0 & 5/21 \\
\bottomrule
\end{tabular}
\end{table}

The LLM policy path remained executable throughout the campaign. For static LLM, one policy is generated before each run. For adaptive LLM-rApp/xApp, the policy layer is queried on the slow 10~s interval, producing 10 policy updates per 100~s run and 210 attempted updates across 21 runs. All adaptive updates were accepted and no update was rejected. The fallback counts in Table~\ref{tab:policy_validation} correspond to controller-level fallback/heuristic events observed during execution, not malformed YAML policies. This distinction is important: the schema-level policy-generation path remained valid, while the xApp-level safety logic still recorded bounded corrective behaviour.

\subsection{Critical-Service Reliability}
\label{sec:results_critical}
\begin{table}[!tb]
\centering
\caption{Critical DC-PRR (\%) by method and density (mean over 7 seeds).}
\label{tab:critical_by_density}
\begin{tabular}{l c c c}
\toprule
\textbf{Method} & \textbf{20 veh.} & \textbf{25 veh.} & \textbf{30 veh.} \\
\midrule
PF                       & 95.1 & 70.2 & 59.8 \\
static expert            & 94.5 & 64.7 & 62.1 \\
balanced expert          & 97.9 & 65.2 & 54.9 \\
heuristic xApp           & 98.6 & 59.6 & 61.1 \\
static LLM               & 90.5 & 69.2 & 63.7 \\
LLM-adaptive             & 84.4 & 62.1 & 65.4 \\
\bottomrule
\end{tabular}
\end{table}

Table~\ref{tab:critical_by_density} reports mean critical DC-PRR by method and density. At 20 vehicles, deterministic and static policies remain in a high-reliability region, while adaptive LLM-rApp/xApp is weaker in mean critical DC-PRR. At 25 vehicles, PF and static LLM are stronger than the adaptive method on this aggregate critical metric, although adaptive LLM-rApp/xApp still remains within the same broad performance band. At 30 vehicles, adaptive LLM-rApp/xApp obtains the highest mean critical DC-PRR among the evaluated methods, reaching 65.4\%, compared with 59.8\% for PF and 63.7\% for static LLM. This supports a high-load directional benefit, while the paired tests below show that it should not be overstated as statistically definitive.

\begin{table}[!tb]
\centering
\caption{ToD UL DC-PRR (\%) by method and density (mean over 7 seeds).}
\label{tab:tod_ul_by_density}
\begin{tabular}{l c c c}
\toprule
\textbf{Method} & \textbf{20 veh.} & \textbf{25 veh.} & \textbf{30 veh.} \\
\midrule
PF                       & 95.1 & 69.4 & 59.1 \\
static expert            & 94.4 & 63.3 & 60.9 \\
balanced expert          & 98.0 & 64.2 & 53.6 \\
heuristic xApp           & 98.7 & 58.2 & 60.1 \\
static LLM               & 90.0 & 68.1 & 62.8 \\
LLM-adaptive             & 83.4 & 60.3 & 65.3 \\
\bottomrule
\end{tabular}
\end{table}

Table~\ref{tab:tod_ul_by_density} isolates ToD UL DC-PRR, the highest-rate critical uplink flow. Adaptive LLM-rApp/xApp again performs best at the highest density, reaching 65.3\% at 30 vehicles. However, it is weaker at 20 and 25 vehicles, especially compared with PF, static LLM, and the deterministic baselines. The result is therefore not that the adaptive policy dominates ToD UL across all regimes, but rather that it becomes more competitive when the network is most congested.

\begin{figure*}[!t]
    \centering
    \includegraphics[width=0.92\textwidth]{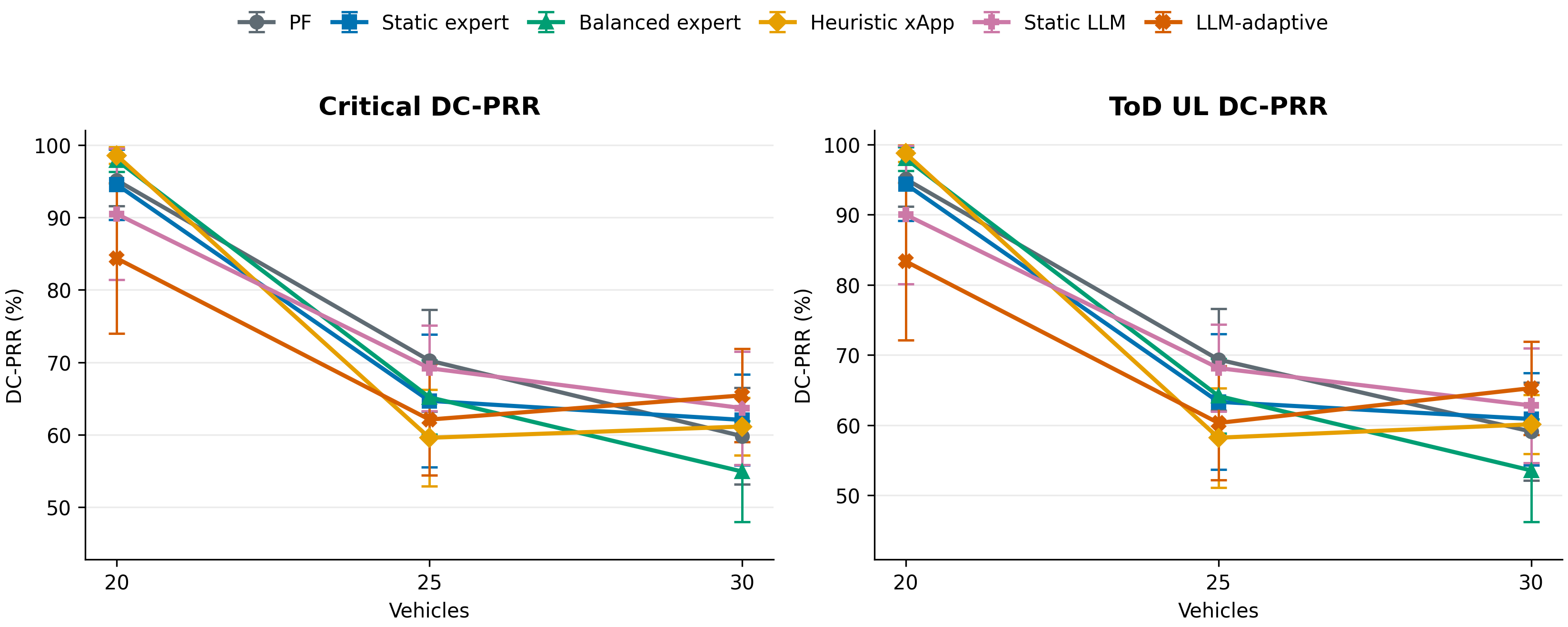}
    \caption{Reliability trends by density. The left panel reports critical DC-PRR, while the right panel isolates ToD UL DC-PRR. Error bars show the standard error of the mean across seven seeds.}
    \label{fig:reliability_trends}
\end{figure*}

\begin{table}[!tb]
\centering
\caption{Awareness UL DC-PRR (\%) by method and density (mean over 7 seeds).}
\label{tab:awareness_ul_by_density}
\begin{tabular}{l c c c}
\toprule
\textbf{Method} & \textbf{20 veh.} & \textbf{25 veh.} & \textbf{30 veh.} \\
\midrule
PF                       & 90.0 & 86.0 & 77.8 \\
static expert            & 91.3 & 81.7 & 83.5 \\
balanced expert          & 92.9 & 82.6 & 79.1 \\
heuristic xApp           & 92.8 & 88.7 & 78.4 \\
static LLM               & 88.8 & 91.0 & 76.5 \\
LLM-adaptive             & 87.9 & 91.1 & 75.7 \\
\bottomrule
\end{tabular}
\end{table}

Awareness UL DC-PRR, shown in Table~\ref{tab:awareness_ul_by_density}, reveals a different pattern. Adaptive LLM-rApp/xApp is strongest at 25 vehicles among the listed methods for this component, but it is not strongest at 20 or 30 vehicles. This confirms that the adaptive policy does not improve all critical directions uniformly. It can protect selected components effectively, but the high-density aggregate result depends on the balance across ToD and Awareness flows.

\begin{figure*}[!t]
    \centering
    \includegraphics[width=0.92\textwidth]{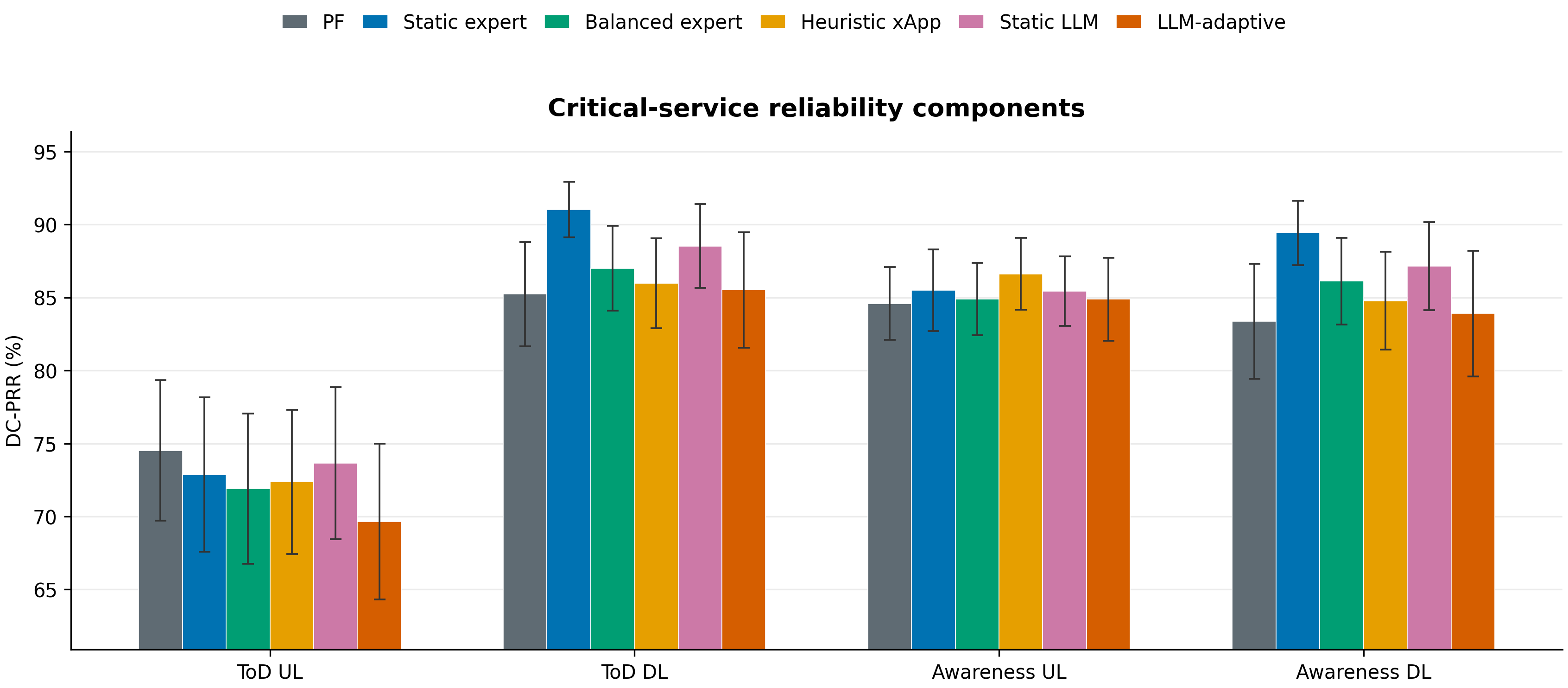}
    \caption{Overall critical-service DC-PRR components across all densities and seeds. Bars show means and error bars show the standard error of the mean across runs.}
    \label{fig:critical_components}
\end{figure*}

\subsection{Deadline Violations and Tail Latency}
\begin{table}[!tb]
\centering
\caption{Critical deadline-violation rate (\%) by method and density, computed as $1-\mathrm{DC\mbox{-}PRR}$ (mean over 7 seeds).}
\label{tab:violations_by_density}
\begin{tabular}{l c c c}
\toprule
\textbf{Method} & \textbf{20 veh.} & \textbf{25 veh.} & \textbf{30 veh.} \\
\midrule
PF                       & 4.9 & 29.8 & 40.2 \\
static expert            & 5.5 & 35.3 & 37.9 \\
balanced expert          & 2.1 & 34.8 & 45.1 \\
heuristic xApp           & 1.4 & 40.4 & 38.9 \\
static LLM               & 9.5 & 30.8 & 36.3 \\
LLM-adaptive             & 15.6 & 37.9 & 34.6 \\
\bottomrule
\end{tabular}
\end{table}

Deadline-violation trends mirror the critical DC-PRR results because the violation rate is computed as one minus the on-time critical packet reception ratio. Adaptive LLM-rApp/xApp has the lowest violation rate at 30 vehicles, but the highest violation rate at 20 and 25 vehicles among the main static/LLM alternatives. This density-dependent reversal is central to the interpretation of the method: the adaptive architecture is useful in stressed regimes, but unnecessary or insufficiently calibrated at lighter operating points.

\begin{table}[!tb]
\centering
\caption{Critical p95 maximum latency (ms) by method and density (mean over 7 seeds).}
\label{tab:latency_by_density}
\begin{tabular}{l c c c}
\toprule
\textbf{Method} & \textbf{20 veh.} & \textbf{25 veh.} & \textbf{30 veh.} \\
\midrule
PF                       & 30.4 & 2038.4 & 2465.9 \\
static expert            & 16.3 & 1725.4 & 1406.2 \\
balanced expert          & 125.6 & 2536.7 & 2195.4 \\
heuristic xApp           & 57.7 & 3252.2 & 1675.0 \\
static LLM               & 16.7 & 2474.6 & 1726.8 \\
LLM-adaptive             & 16.1 & 1329.3 & 3825.8 \\
\bottomrule
\end{tabular}
\end{table}

Critical p95 maximum latency is high variance under congestion. At 25 vehicles, adaptive LLM-rApp/xApp gives the lowest mean critical p95 maximum latency among all methods, at 1329.3~ms, compared with 2038.4~ms for PF and 3252.2~ms for heuristic xApp. At 30 vehicles, however, it has the worst value, showing that the high-density reliability gain comes with a tail-latency cost. This is why latency claims are reported as operating-point evidence rather than a universal improvement.

\subsection{Background-Service Preservation}
\begin{table}[!tb]
\centering
\caption{Background throughput (Mbps) by method and density (mean over 7 seeds).}
\label{tab:background_by_density}
\begin{tabular}{l c c c}
\toprule
\textbf{Method} & \textbf{20 veh.} & \textbf{25 veh.} & \textbf{30 veh.} \\
\midrule
PF                       & 16.272 & 12.930 & 11.118 \\
static expert            & 16.359 & 13.116 & 12.936 \\
balanced expert          & 16.175 & 12.398 & 11.840 \\
heuristic xApp           & 16.224 & 11.915 & 12.399 \\
static LLM               & 16.425 & 13.081 & 11.973 \\
LLM-adaptive             & 16.425 & 14.510 & 9.537 \\
\bottomrule
\end{tabular}
\end{table}

Table~\ref{tab:background_by_density} reports background throughput. Adaptive LLM-rApp/xApp preserves the highest background throughput at 25 vehicles, reaching 14.510~Mbps, but it is weaker at 30 vehicles. This is one of the clearest medium-density strengths of the adaptive method: it improves the throughput side of the reliability--throughput trade-off at 25 vehicles. At 30 vehicles, static expert and heuristic xApp preserve substantially more background throughput, indicating that the adaptive method prioritizes critical traffic more aggressively under high congestion.

\subsection{Benefit of xApp-Level Execution}
\label{sec:results_static_vs_adaptive}
Comparing static LLM against adaptive LLM-rApp/xApp isolates the contribution of periodic policy refresh and deterministic xApp-level execution. The adaptive method is not uniformly better than the static LLM policy: static LLM is stronger in aggregate critical DC-PRR and background throughput. However, adaptive LLM-rApp/xApp is better in several important operating points, particularly critical DC-PRR and ToD UL DC-PRR at 30 vehicles, background throughput at 25 vehicles, and critical p95 maximum latency at 25 vehicles. These results suggest that xApp-level execution is most useful in specific traffic-density regimes rather than as a universal replacement for well-calibrated static policies.

\begin{figure*}[!t]
    \centering
    \includegraphics[width=0.92\textwidth]{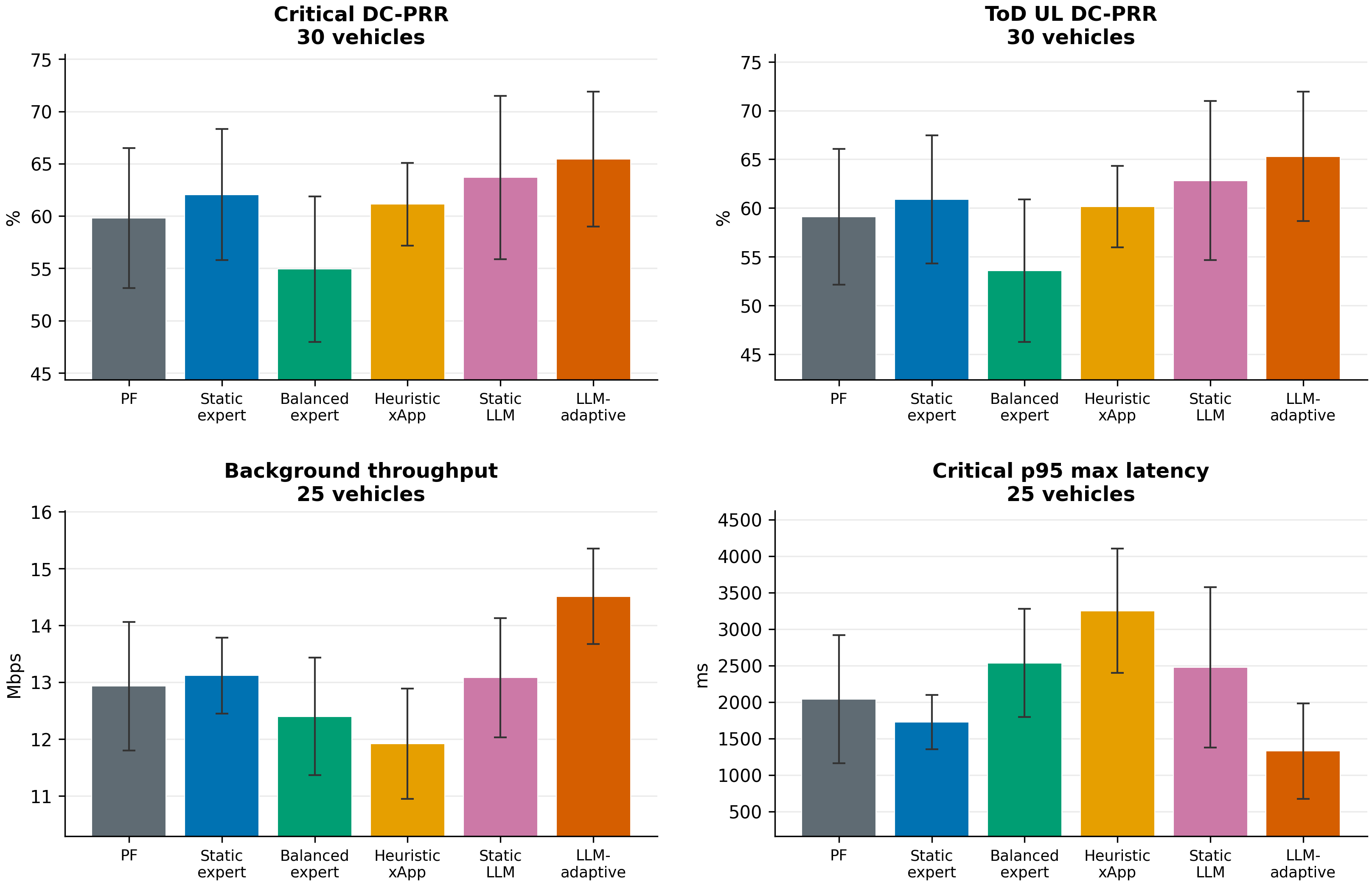}
    \caption{Selected operating points where adaptive LLM-rApp/xApp is strongest or especially competitive: high-density critical reliability, high-density ToD UL reliability, medium-density background throughput, and medium-density critical p95 maximum latency.}
    \label{fig:adaptive_strengths}
\end{figure*}

\subsection{Control Overhead and Practicality}
\label{sec:results_overhead}
\begin{table}[!tb]
\centering
\caption{Control overhead and practicality, LLM-based methods (mean across 21 runs unless noted).}
\label{tab:control_overhead}
\begin{tabular}{l c c}
\toprule
\textbf{Metric} & \textbf{static LLM} & \textbf{LLM-adaptive} \\
\midrule
LLM generation latency, mean (s) & --\textsuperscript{a} & 8.36 \\
LLM generation latency, max observed (s) & --\textsuperscript{a} & 16.80 \\
Policy generations per run & 1 & 10 \\
Policy updates accepted per run & 1 & 10 \\
Policy updates rejected per run & 0 & 0 \\
Fallback invocations per run & 0.29 & 0.24 \\
Rollback rows per run & -- & 229.3 \\
Directional UL interventions per run & -- & 46.2 \\
Directional DL interventions per run & -- & 37.8 \\
xApp mean step latency (ms) & 0.014 & 0.028 \\
\bottomrule
\end{tabular}
\begin{flushleft}
\footnotesize\textsuperscript{a}static LLM issues a single generation call before the run; its per-call latency was not separately logged.
\end{flushleft}
\end{table}

The measured control overhead confirms the intended timescale separation. The LLM operates on the slow policy-generation interval and remains outside the 100~ms xApp loop. The reported LLM latency is wall-clock time, whereas the 10~s policy refresh interval is simulation time; therefore, the LLM generation latency should not be interpreted as a real-time scheduler latency. The deterministic xApp-like controller remains lightweight, with sub-millisecond mean execution overhead in the logged runs.

\subsection{Aggregate Comparison}
\label{sec:results_aggregate}
\begin{table*}[!t]
\centering
\caption{Aggregate performance comparison across all 126 runs (mean over 3 densities $\times$ 7 seeds = 21 runs per method).}
\label{tab:aggregate_results}
\begin{tabular}{l c c c c c c c}
\toprule
\textbf{Method} &
\textbf{Crit DC-PRR} &
\textbf{ToD UL} &
\textbf{AIC UL} &
\textbf{Viol. Rate} &
\textbf{Bg. Mbps} &
\textbf{Mean Delay} &
\textbf{p95 Max} \\
\midrule
PF                       & 75.1 & 74.5 & 84.6 & 24.9 & 13.440 & 79.4 & 1511.6 \\
static expert            & 73.8 & 72.9 & 85.5 & 26.2 & 14.137 & 46.6 & 1049.3 \\
balanced expert          & 72.7 & 71.9 & 84.9 & 27.3 & 13.471 & 80.3 & 1619.2 \\
heuristic xApp           & 73.1 & 72.4 & 86.6 & 26.9 & 13.513 & 88.2 & 1661.6 \\
static LLM               & 74.4 & 73.7 & 85.4 & 25.6 & 13.826 & 75.3 & 1406.0 \\
LLM-adaptive             & 70.6 & 69.7 & 84.9 & 29.4 & 13.490 & 96.5 & 1723.7 \\
\bottomrule
\end{tabular}
\end{table*}

Table~\ref{tab:aggregate_results} aggregates the headline metrics across all densities and seeds. PF has the highest aggregate critical DC-PRR in this seven-seed campaign, followed by static LLM, static expert, heuristic xApp, balanced expert, and adaptive LLM-rApp/xApp. Therefore, the adaptive method is not the best aggregate critical-reliability method. Its aggregate background throughput is competitive but not the strongest, and its aggregate p95 maximum latency is also not the best. The aggregate view therefore supports the conservative framing of the paper: Agentic-V2X demonstrates a validated and competitive architecture for LLM-assisted policy generation, but it does not establish universal scheduler dominance.

\begin{figure*}[!t]
    \centering
    \includegraphics[width=0.92\textwidth]{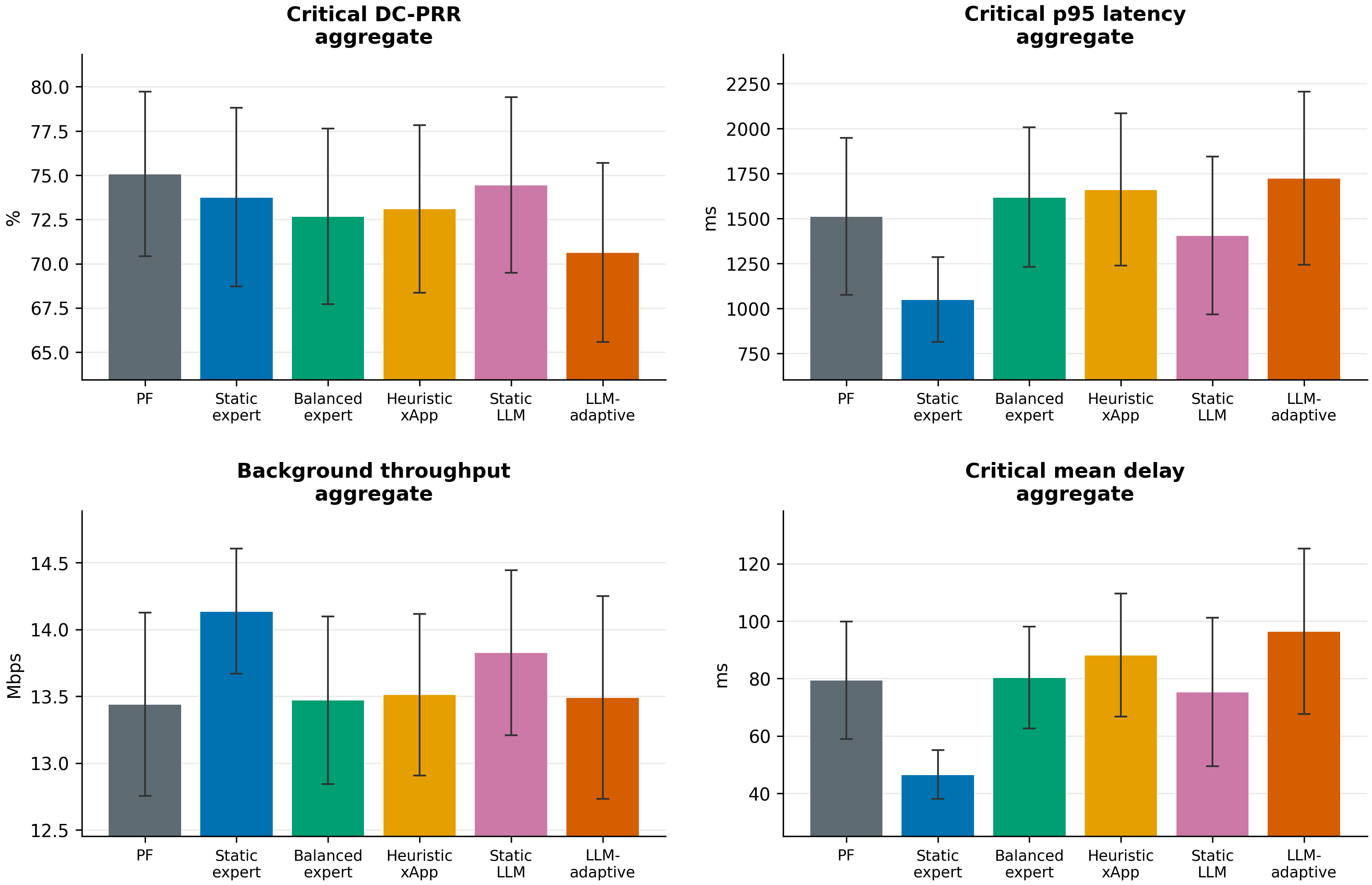}
    \caption{Aggregate performance trade-offs across all densities and seeds. Higher is better for critical DC-PRR and background throughput, while lower is better for deadline-violation rate and latency metrics.}
    \label{fig:aggregate_tradeoffs}
\end{figure*}

\subsection{Paired Statistical Analysis}
To avoid treating repeated simulation runs as independent observations, we performed paired statistical analysis over matched density/seed combinations. For each comparison, paired differences are computed as adaptive LLM-rApp/xApp minus the baseline on the same density and seed. The aggregate paired tests therefore use 21 matched pairs, while density-specific tests use seven matched pairs. Positive differences indicate improvements for DC-PRR and throughput metrics, whereas negative differences indicate improvements for latency metrics. For each comparison, we report the mean paired difference, median paired difference, 95\% paired bootstrap confidence interval for the mean difference, two-sided Wilcoxon signed-rank $p$-value, and paired win rate.

\begin{table*}[!t]
\centering
\caption{Aggregate paired statistical comparison for critical DC-PRR. Differences are LLM-adaptive minus baseline over matched density/seed pairs; positive values favour LLM-adaptive.}
\label{tab:paired_stats_critical}
\begin{tabular}{l c c c c c c}
\toprule
\textbf{Baseline} & \textbf{$n$} & \textbf{Mean diff. (pp)} & \textbf{Median diff. (pp)} & \textbf{95\% bootstrap CI (pp)} & \textbf{Wilcoxon $p$} & \textbf{Win rate} \\
\midrule
PF & 21 & -4.4 & +0.1 & [-14.4, +4.2] & 0.759 & 61.9\% \\
heuristic xApp & 21 & -2.5 & +0.0 & [-12.3, +5.7] & 0.973 & 52.4\% \\
static LLM & 21 & -3.8 & +0.0 & [-10.9, +2.9] & 0.585 & 52.4\% \\
static expert & 21 & -3.1 & -0.3 & [-12.0, +4.2] & 0.683 & 42.9\% \\
\bottomrule
\end{tabular}
\end{table*}

Table~\ref{tab:paired_stats_critical} shows that adaptive LLM-rApp/xApp is not the best aggregate method for critical DC-PRR. The mean paired difference is negative against PF, heuristic xApp, static LLM, and static expert, and all bootstrap confidence intervals include zero. The Wilcoxon signed-rank tests are also not significant. Interestingly, the paired win rate against PF is 61.9\%, even though the mean difference is negative, which indicates that a small number of large negative cases dominate the aggregate mean. This supports a competitive but not dominant interpretation.

\begin{figure}[!tb]
    \centering
    \includegraphics[width=\linewidth]{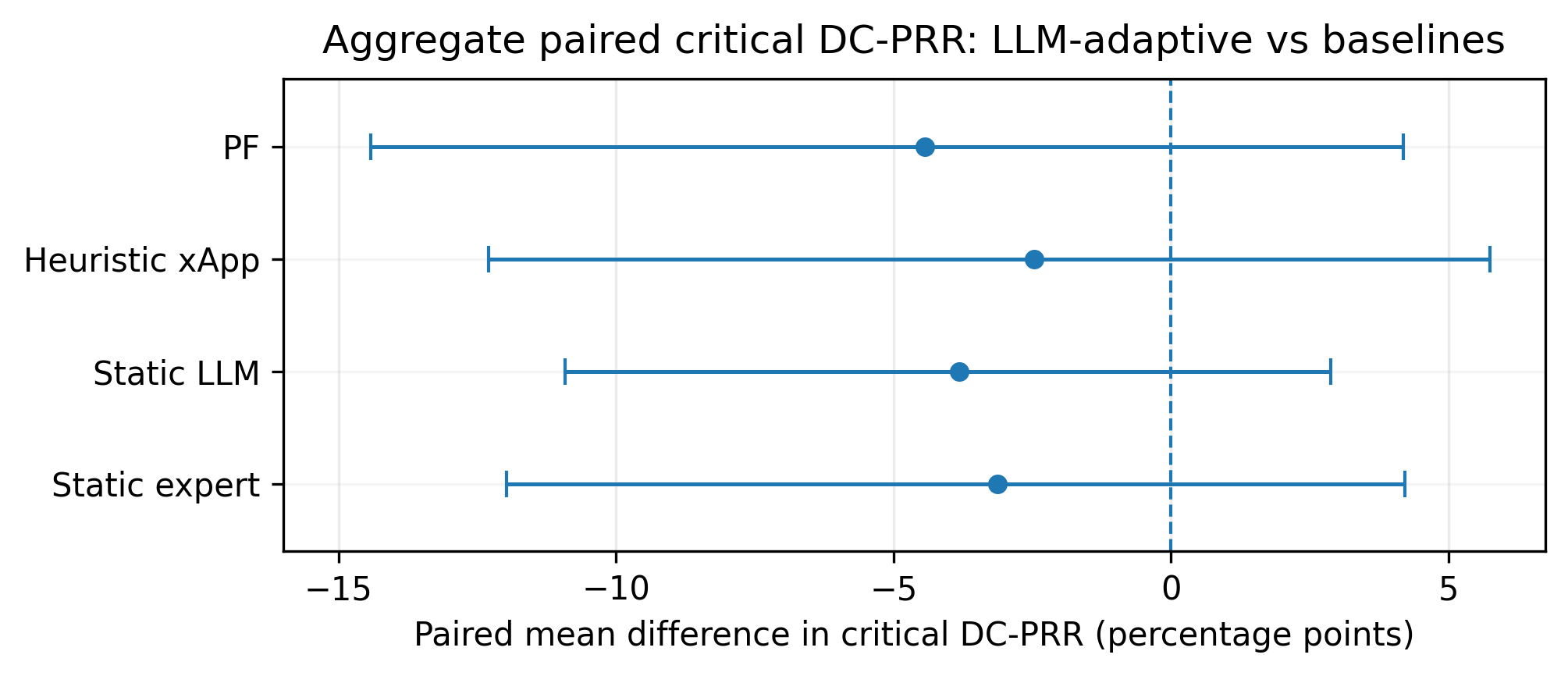}
    \caption{Aggregate paired critical DC-PRR differences for adaptive LLM-rApp/xApp against selected baselines over matched density/seed pairs.}
    \label{fig:paired_critical}
\end{figure}

\begin{table*}[!t]
\centering
\caption{Aggregate paired statistics for the primary reliability metric and key secondary metrics. Differences are adaptive LLM-rApp/xApp minus baseline over 21 matched density/seed pairs.}
\label{tab:paired_stats_aggregate_metrics}
\scriptsize
\resizebox{\textwidth}{!}{%
\begin{tabular}{l l c c c c c c c}
\toprule
\textbf{Metric} & \textbf{Baseline} & \textbf{Unit} & \textbf{$n$} & \textbf{Mean diff.} & \textbf{Median diff.} & \textbf{95\% bootstrap CI} & \textbf{Wilcoxon $p$} & \textbf{Win rate} \\
\midrule
Critical DC-PRR & PF & pp & 21 & -4.4 & +0.1 & [-14.4, +4.2] & 0.759 & 61.9\% \\
 & heuristic xApp & pp & 21 & -2.5 & +0.0 & [-12.3, +5.7] & 0.973 & 52.4\% \\
 & static LLM & pp & 21 & -3.8 & +0.0 & [-10.9, +2.9] & 0.585 & 52.4\% \\
 & static expert & pp & 21 & -3.1 & -0.3 & [-12.0, +4.2] & 0.683 & 42.9\% \\
Critical p95 max & PF & ms & 21 & +212.2 & +0.0 & [-524.0, +997.3] & 0.812 & 47.6\% \\
 & heuristic xApp & ms & 21 & +62.1 & +0.3 & [-921.3, +1106.0] & 0.658 & 38.1\% \\
 & static LLM & ms & 21 & +317.7 & +1.3 & [-879.3, +1438.4] & 0.393 & 42.9\% \\
 & static expert & ms & 21 & +674.4 & +0.1 & [+2.2, +1431.0] & 0.473 & 47.6\% \\
Background throughput & PF & Mbps & 21 & +0.050 & -0.000 & [-0.877, +1.083] & 0.794 & 47.6\% \\
 & heuristic xApp & Mbps & 21 & -0.022 & +0.000 & [-1.424, +1.289] & 0.865 & 52.4\% \\
 & static LLM & Mbps & 21 & -0.336 & -0.000 & [-1.711, +1.043] & 0.502 & 38.1\% \\
 & static expert & Mbps & 21 & -0.646 & -0.000 & [-1.797, +0.449] & 0.391 & 38.1\% \\
\bottomrule
\end{tabular}%
}
\end{table*}

Table~\ref{tab:paired_stats_aggregate_metrics} extends the aggregate analysis to the main secondary metrics. The adaptive method is not statistically superior in aggregate critical reliability. For latency and background throughput, the signs of the mean paired differences are mixed across baselines, and most confidence intervals include zero. We therefore interpret these metrics as evidence of trade-off behaviour rather than universal dominance.

\begin{table*}[!t]
\centering
\caption{Paired statistical checks for the main operating-point claims. Differences are adaptive LLM-rApp/xApp minus the baseline on the same density and seed. Positive values favour adaptive LLM for DC-PRR and throughput; negative values favour adaptive LLM for latency.}
\label{tab:paired_claim_checks}
\scriptsize
\resizebox{\textwidth}{!}{%
\begin{tabular}{l l c c c c c c c}
\toprule
\textbf{Claim / metric} & \textbf{Baseline} & \textbf{Unit} & \textbf{$n$} & \textbf{Mean diff.} & \textbf{Median diff.} & \textbf{95\% bootstrap CI} & \textbf{Wilcoxon $p$} & \textbf{Win rate} \\
\midrule
Critical DC-PRR, 30 veh. & PF & pp & 7 & +5.6 & +7.3 & [-1.3, +11.6] & 0.156 & 71.4\% \\
 & heuristic xApp & pp & 7 & +4.3 & +8.6 & [-6.3, +14.9] & 0.578 & 57.1\% \\
 & static LLM & pp & 7 & +1.8 & +0.5 & [-6.3, +10.5] & 0.688 & 71.4\% \\
 & static expert & pp & 7 & +3.4 & -0.9 & [-6.7, +13.7] & 0.688 & 42.9\% \\
Critical p95 max, 25 veh. & PF & ms & 7 & -709.1 & -1104.4 & [-2007.1, +676.5] & 0.469 & 57.1\% \\
 & heuristic xApp & ms & 7 & -1922.9 & -2370.7 & [-3413.2, -408.1] & 0.109 & 71.4\% \\
 & static LLM & ms & 7 & -1145.3 & -426.1 & [-3858.2, +1621.0] & 0.469 & 57.1\% \\
 & static expert & ms & 7 & -396.1 & -671.4 & [-1004.3, +257.6] & 0.375 & 71.4\% \\
Background throughput, 25 veh. & PF & Mbps & 7 & +1.580 & +0.749 & [-0.202, +3.677] & 0.375 & 57.1\% \\
 & heuristic xApp & Mbps & 7 & +2.595 & +2.248 & [+0.824, +4.419] & 0.047 & 85.7\% \\
 & static LLM & Mbps & 7 & +1.429 & +1.861 & [-1.503, +4.126] & 0.469 & 57.1\% \\
 & static expert & Mbps & 7 & +1.394 & +0.642 & [+0.180, +2.842] & 0.219 & 57.1\% \\
Critical DC-PRR, 20 veh. & PF & pp & 7 & -10.8 & +0.0 & [-34.7, +8.8] & 0.938 & 71.4\% \\
 & heuristic xApp & pp & 7 & -14.2 & -0.3 & [-34.7, +2.0] & 0.219 & 28.6\% \\
 & static LLM & pp & 7 & -6.1 & -0.1 & [-17.1, +0.1] & 0.375 & 42.9\% \\
 & static expert & pp & 7 & -10.2 & -0.3 & [-30.0, +0.5] & 0.297 & 28.6\% \\
Aggregate critical DC-PRR & PF & pp & 21 & -4.4 & +0.1 & [-14.4, +4.2] & 0.759 & 61.9\% \\
 & heuristic xApp & pp & 21 & -2.5 & +0.0 & [-12.3, +5.7] & 0.973 & 52.4\% \\
 & static LLM & pp & 21 & -3.8 & +0.0 & [-10.9, +2.9] & 0.585 & 52.4\% \\
 & static expert & pp & 21 & -3.1 & -0.3 & [-12.0, +4.2] & 0.683 & 42.9\% \\
\bottomrule
\end{tabular}%
}
\end{table*}

Table~\ref{tab:paired_claim_checks} reports the density-specific claim checks. At 30 vehicles, adaptive LLM-rApp/xApp improves mean critical DC-PRR over PF by 5.6 percentage points, with a 71.4\% paired win rate, but the Wilcoxon test does not reach significance. At 25 vehicles, adaptive LLM-rApp/xApp reduces critical p95 maximum latency relative to all four selected baselines and improves background throughput most clearly against heuristic xApp, where the paired Wilcoxon result is favourable ($p=0.047$). At 20 vehicles, adaptive LLM-rApp/xApp is directionally worse in mean critical DC-PRR, especially against heuristic xApp and static expert. These checks show that the adaptive architecture has density-dependent strengths, but the aggregate claim must remain conservative.

\subsection{Summary of Findings}
Three main findings come out of this work. First, the LLM policy path remained executable across the seven-seed campaign, with all adaptive policy updates accepted and no rejected updates. Second, adaptive LLM-rApp/xApp is not the best aggregate reliability method, but it shows useful density-specific strengths: critical DC-PRR and ToD UL DC-PRR at 30 vehicles, background throughput at 25 vehicles, and critical p95 maximum latency at 25 vehicles. Third, paired statistical analysis over matched density/seed pairs confirms that the strongest defensible conclusion is architectural competitiveness rather than universal dominance.

\section{Discussion}

\subsection{LLM as a Real-Time Scheduler}
The architecture deliberately keeps the LLM out of the scheduling loop. Three properties of LLM inference make it unsuitable for near-real-time scheduling: latency at a multi-second wall-clock scale in this configuration, non-deterministic outputs that complicate reproducibility and verification, and the absence of native guarantees of validity, boundedness, or safety. By confining the LLM to slow policy generation and assigning all fast actuation to a deterministic, bounded controller, the design uses the LLM for reasoning over objectives, priorities, and trade-offs without exposing the radio control loop to its weaknesses. The measured LLM generation latency supports this design choice, and it must be interpreted separately from the 10~s simulation-time policy interval.

\subsection{Small LLMs as rApp-Inspired Policy Creators}
\label{sec:disc_validity}
The validity results show that \texttt{qwen2.5:7b}, prompted with the constrained weight-policy schema used in this study, produced executable, in-bounds policies across all static and adaptive policy-generation/update attempts. This supports the claim that a small local model can act as an rApp-inspired policy creator for this constrained task. We are careful, however, not to over-generalise this finding: a high validity rate on a deliberately constrained schema does not imply that arbitrary LLM outputs are safe. The validation, repair, fallback, and xApp-level bounding layers remain essential to the architecture.

\subsection{Trade-off Interpretation}
\label{sec:disc_tradeoff}
The results show a more nuanced picture than a simple ranking. Adaptive LLM-rApp/xApp is strongest in selected operating points, especially at 30 vehicles for critical reliability and ToD UL reliability, and at 25 vehicles for background throughput and critical p95 maximum latency. In aggregate, however, it remains below the strongest methods on critical DC-PRR. We therefore interpret the adaptive method as a competitive policy-generation architecture with useful density-specific strengths, not as a universally best scheduler.

Two factors plausibly explain the mixed outcome. First, the 10~s policy-refresh period is coarse relative to how quickly contention builds and dissipates as vehicles move through the Manhattan grid. Second, the xApp adaptation rules are deliberately conservative to avoid oscillation and safety violations, which may limit their ability to correct a suboptimal policy quickly in the highest-density case. Faster refresh, richer telemetry summaries, or more targeted critical-flow balancing may widen the benefit.

\subsection{Role of the xApp}
Because the deterministic xApp-like controller is shared in design between heuristic xApp and adaptive LLM-rApp/xApp, their differences reflect the policy bounds and baseline weights supplied by the LLM versus the hand-coded heuristic rule set. The adaptive method is strongest at specific load points rather than uniformly across the full campaign. This suggests that the xApp execution layer is useful, but that the policy-generation layer still requires better calibration for low-load operation and for balancing multiple critical-flow directions under heavy congestion.

\subsection{Safety and Validation}
The validation, repair, and fallback layer is what makes delegating policy creation to a stochastic model defensible. Because every policy is checked and bounded before execution, and because the worst case degrades to a known deterministic policy, a malformed or unsafe LLM output cannot directly reach the scheduler. The seven-seed campaign did not produce rejected policy updates, but the controller did record bounded fallback/rollback behaviour during execution, showing the importance of runtime safety checks in addition to schema validation.

\subsection{Failure Cases}
\label{sec:disc_failure}
The main failure case for adaptive LLM-rApp/xApp appears at low density and in aggregate performance. At 20 vehicles, the adaptive method is directionally worse on critical DC-PRR, suggesting that adaptation may be unnecessary or even harmful when the network is not strongly congested. At 30 vehicles, the adaptive method improves critical reliability and ToD UL reliability, but this comes with poor tail latency and weaker background throughput. These two failure modes show that valid LLM-generated policies are not automatically well calibrated; the schema and validator ensure safety and executability, but performance still depends on the content and timing of the generated policy.

\section{Limitations}
This study has several limitations. First, it is entirely simulation-based. Results from ns-3/5G-LENA/SUMO may not transfer directly to hardware. The rApp/xApp terminology is conceptual: we do not implement the A1/E2 interfaces, a real Non-RT or Near-RT RIC, or a conformant rApp/xApp, and the generated policy is a validated structured policy in an rApp-inspired framework, not a certified rApp. The evaluation uses a single 5G NR cell, three densities, and a single fixed traffic regime determined by the service models. We study one representative small local model (\texttt{qwen2.5:7b}). The work is not a model benchmark and does not claim this model is optimal. The control action space is limited to per-service scheduler weights mapped to per-UE UL/DL weights over a PF-based scheduler, and the LLM update cadence (10~s) and control period (100~ms) are fixed. The campaign uses seven seeds per density-method cell, which provides a cleaner paired comparison than a minimal pilot design but is still modest for density-specific hypothesis testing. Aggregate paired tests use 21 matched density/seed pairs, while density-specific tests use only seven matched pairs; therefore small numerical differences and high-variance tail-latency metrics should be interpreted cautiously. Finally, results may depend on the prompt and schema design, while robustness to alternative prompts, schemas, and models is left to future work.

\section{Conclusion}
We presented Agentic-V2X, a simulation-based framework in which a small, locally deployable language model acts as an rApp-inspired policy agent that generates structured, validated scheduler-configuration policies for deadline-aware 5G NR V2X traffic, while a deterministic xApp-like controller executes those policies through ns3-ai at a 100~ms cadence. Across the full 126-run campaign (3 densities $\times$ 7 seeds $\times$ 6 methods), the small local LLM (\texttt{qwen2.5:7b}) produced executable adaptive policies with all 210 policy updates accepted and no rejected updates. On network performance, adaptive LLM-rApp/xApp was strongest in several important operating points, including critical DC-PRR and ToD UL DC-PRR at 30 vehicles, background throughput at 25 vehicles, and critical p95 maximum latency at 25 vehicles. At the same time, it remained below the strongest methods on aggregate critical reliability. The main lesson is therefore balanced: small LLM agents can safely generate executable V2X scheduler policies and can be competitive in specific density regimes, but they should be treated as policy assistants within a validated hierarchical control architecture rather than as replacements for well-calibrated deterministic control.

\section{Future Work}
Future work includes integration with a real O-RAN Non-RT/Near-RT RIC and A1/E2 interfaces, multi-cell scenarios when handover becomes available in 5G-LENA ns3, a richer policy schema and broader model and prompt studies to test whether the observed policy validity holds under harder conditions, faster or telemetry-triggered policy refresh to widen the density-specific gains identified here, and targeted prompt or training interventions to improve low-load stability and high-density critical-flow balance without sacrificing ToD uplink protection.

\bibliographystyle{IEEEtran}
\bibliography{references}

\clearpage
\onecolumn
\appendix
\section{LLM-rApp Policy Generation Artifacts}
\label{app:llm-rapp-artifacts}

This appendix reports the main artifacts used by the rApp-inspired LLM policy generator. It includes the prompt structure, the constrained YAML policy interface accepted by the deterministic xApp-like validator, and one complete policy generated and accepted during the ns-3 campaign. These artifacts are included to make the LLM-to-controller interface auditable and reproducible. In the implementation, the runtime prompt is constructed by concatenating the system prompt and the user prompt after substituting the scenario identifier, telemetry summary, objective identifier, model name, and policy identifier.

\subsection{Prompt Structure}
\label{app:llm-prompt-structure}

The LLM is not used as a real-time scheduler. Instead, it generates a bounded service-level policy that is validated and then executed by the deterministic xApp-like controller. The system prompt constrains the model to return YAML only, restricts the admissible services, metrics, actions, and scheduler-weight ranges, and specifies the safety requirements that must hold before a policy can be accepted.
\\
\subsubsection{System Prompt}
\label{app:system-prompt}

\noindent\textbf{Listing A.1: System prompt used by the rApp-inspired LLM policy generator.}

\begin{lstlisting}[style=agenticv2x]
You are a 5G NR V2X rApp policy generator.

Your task is to generate one structured scheduler policy for a deterministic xApp controller.

You are not the online scheduler.
You do not output immediate runtime actions.
You only output a policy that the xApp can validate and execute.

Return only valid YAML.
Do not include explanations.
Do not include Markdown fences.
Do not include comments.
Do not include text before or after the YAML.

# V2X service criticality and latency requirements

ToD (Time-critical Operational Data) and Awareness (AIC) are safety-of-life services.
They must always receive high scheduler priority.

ToD carries operator command downlink and sensor/video uplink.
- The DL direction is the tightest control path.
- Violation of the DL deadline directly impairs safety-critical vehicle control loops.
- ToD must receive the highest baseline weight and must never be starved under congestion.

Awareness (AIC) carries cooperative awareness messages between vehicles.
- Deadline: 100 ms p95 in both UL and DL directions.
- Violation impairs situational awareness and collision avoidance.
- Awareness must be protected under congestion.
- Awareness must not be sacrificed to free bandwidth for HDMap or Sensor.

HDMap carries high-definition map data.
- This is a background throughput service.
- Degradation up to 35% is tolerable.
- HDMap must not receive zero weight but can be deprioritised under congestion.

Sensor (RTSA) carries real-time sensor data at low rates.
- This is a background service.
- It must not receive zero weight but can be trimmed under congestion.

Under congestion, the correct priority order is:
ToD > Awareness > HDMap approximately Sensor.

# Allowed services

- ToD
- Awareness
- HDMap
- Sensor

# Allowed telemetry metrics

- tod_p95_latency_ms
- tod_dc_prr
- tod_violation_rate
- awareness_p95_latency_ms
- awareness_dc_prr
- awareness_violation_rate
- hdmap_throughput_degradation
- sensor_throughput_degradation
- congestion_ratio

# Allowed actions

- set_weight
- increase_weight
- decrease_weight

Allowed scheduler weights are integers from 1 to 10.

Telemetry values use the following numeric ranges:
- dc_prr metrics are ratios from 0.0 to 1.0.
- violation_rate metrics are ratios from 0.0 to 1.0.
- throughput_degradation metrics are ratios from 0.0 to 1.0, not percentages.
- congestion_ratio is a ratio from 0.0 to 1.0.
- latency metrics are expressed in milliseconds.

The policy must follow this exact YAML schema:

policy_id: string
description: string
scenario_id: string
objective_id: string
baseline_weights:
  ToD: integer
  Awareness: integer
  HDMap: integer
  Sensor: integer
bounds:
  ToD: [integer, integer]
  Awareness: [integer, integer]
  HDMap: [integer, integer]
  Sensor: [integer, integer]
safety_constraints:
  min_critical_weight: integer
  max_background_degradation: float
  forbid_zero_weight: true
  critical_services:
    - ToD
    - Awareness
xapp_rules:
  - name: string
    condition:
      metric: one allowed telemetry metric
      operator: one of ">", ">=", "<", "<="
      value: number
    action:
      service: one allowed service
      type: one allowed action
      value: integer
fallback_policy: static_expert
runtime_safety:
  rollback_enabled: true
  degradation_windows: integer
  monitored_metrics:
    - critical_violation_rate
    - critical_dc_prr
    - critical_p95_latency_ms
  directional_guard_enabled: true
  max_ul_dl_dc_prr_gap: float
metadata:
  source: llm_rapp
  model_name: string
  generation_time: null

# Safety requirements

- ToD and Awareness are critical services. Their baseline weights must be high.
- ToD must have the highest baseline weight.
- ToD and Awareness must not drop below the minimum critical weight.
- HDMap and Sensor must not receive zero weight.
- Baseline weights must be within their bounds.
- Bounds must be within 1 and 10.
- Lower bounds must not exceed upper bounds.
- Do not invent services.
- Do not invent telemetry metrics.
- Do not invent action types.
- Do not use nested conditions.
- Do not use mathematical expressions.
- Do not output Python code.
- If runtime_safety is included, degradation_windows must be 3 and max_ul_dl_dc_prr_gap must be 0.05.

A good policy is conservative, valid, and executable.
\end{lstlisting}

\subsubsection{User Prompt Template}
\label{app:user-prompt-template}

The user prompt provides the scenario-specific telemetry and identifiers. The telemetry block is generated by the simulator-side logging pipeline and summarises the recent network state available to the slow policy-generation loop.

\noindent\textbf{Listing A.2: User prompt template used to instantiate a scenario-specific LLM-rApp policy request.}

\begin{lstlisting}[style=agenticv2x]
Create one rApp scheduler policy for the scenario and objective below.
The deterministic xApp will execute the policy at runtime using only validated simple rules, bounds, and safety constraints.

{telemetry_summary}

Use these identifiers exactly:
scenario_id: {scenario_id}
objective_id: {objective_id}
model_name: {model_name}

Use this policy_id format:
policy_id: {policy_id}

Return only YAML. Do not use Markdown fences.
Use ratios, not percentages, for DC-PRR, violation-rate, throughput-degradation, and congestion thresholds.

A valid conservative example shape is:

policy_id: {policy_id}
description: Conservative rApp policy for V2X scheduler weights
scenario_id: {scenario_id}
objective_id: {objective_id}
baseline_weights:
  ToD: 7
  Awareness: 6
  HDMap: 3
  Sensor: 2
bounds:
  ToD: [5, 10]
  Awareness: [5, 10]
  HDMap: [1, 8]
  Sensor: [1, 8]
safety_constraints:
  min_critical_weight: 5
  max_background_degradation: 0.35
  forbid_zero_weight: true
  critical_services:
    - ToD
    - Awareness
xapp_rules:
  - name: tod_latency_guard
    condition:
      metric: tod_p95_latency_ms
      operator: ">"
      value: 15
    action:
      service: ToD
      type: increase_weight
      value: 1
  - name: awareness_violation_guard
    condition:
      metric: awareness_violation_rate
      operator: ">"
      value: 0.02
    action:
      service: Awareness
      type: increase_weight
      value: 1
  - name: congestion_sensor_trim
    condition:
      metric: congestion_ratio
      operator: ">"
      value: 0.8
    action:
      service: Sensor
      type: decrease_weight
      value: 1
fallback_policy: static_expert
runtime_safety:
  rollback_enabled: true
  degradation_windows: 3
  monitored_metrics:
    - critical_violation_rate
    - critical_dc_prr
    - critical_p95_latency_ms
  directional_guard_enabled: true
  max_ul_dl_dc_prr_gap: 0.05
metadata:
  source: llm_rapp
  model_name: {model_name}
  generation_time: null
\end{lstlisting}

\subsection{Validated YAML Policy Interface}
\label{app:yaml-policy-schema}

The deterministic validator accepts only the fields and values listed in Listing~A.3. Policies that violate the schema are either repaired deterministically, when the error is recoverable, or rejected and replaced by the fallback policy. The runtime-safety block exposes the xApp shield parameters used for rollback and critical UL/DL balancing. This design ensures that the stochastic LLM output is never applied directly to the scheduler.

\noindent\textbf{Listing A.3: Validated YAML policy interface accepted by the xApp-like controller.}

\begin{lstlisting}[style=agenticv2x]
Required top-level fields:
  policy_id: string
  description: string
  scenario_id: string
  objective_id: string
  baseline_weights: map[service -> integer]
  bounds: map[service -> [integer, integer]]
  safety_constraints: map
  xapp_rules: list[rule]
  fallback_policy: static_expert
  runtime_safety: map
  metadata: map

Allowed services:
  ToD
  Awareness
  HDMap
  Sensor

Weight domain:
  minimum: 1
  maximum: 10

Default validated bounds:
  ToD: [5, 10]
  Awareness: [5, 10]
  HDMap: [1, 10]
  Sensor: [1, 10]

Default safety constraints:
  min_critical_weight: 5
  max_background_degradation: 0.35
  forbid_zero_weight: true
  critical_services:
    - ToD
    - Awareness

Supported telemetry metrics:
  tod_p95_latency_ms
  tod_dc_prr
  tod_violation_rate
  awareness_p95_latency_ms
  awareness_dc_prr
  awareness_violation_rate
  hdmap_throughput_degradation
  sensor_throughput_degradation
  congestion_ratio

Supported operators:
  >
  >=
  <
  <=

Supported action types:
  set_weight
  increase_weight
  decrease_weight

Supported metadata source values:
  llm_rapp
  manual
  baseline
  repaired

Runtime safety shape:
  rollback_enabled: true
  degradation_windows: 3
  monitored_metrics:
    - critical_violation_rate
    - critical_dc_prr
    - critical_p95_latency_ms
  directional_guard_enabled: true
  max_ul_dl_dc_prr_gap: 0.05

Rule shape:
  name: string
  condition:
    metric: supported telemetry metric
    operator: supported operator
    value: numeric threshold
  action:
    service: supported service
    type: supported action type
    value: integer
\end{lstlisting}

\subsection{Example Accepted LLM-Generated Policy}
\label{app:generated-policy-example}

Listing~A.4 reports one complete policy generated by \texttt{qwen2.5:7b}, validated by the policy checker, and accepted during the ns-3 campaign. The example corresponds to the first rApp update of the \texttt{real\_100s\_llm\_rapp\_xapp\_20veh\_seed6} experiment. It illustrates the bounded service-level configuration passed from the slow LLM-rApp layer to the deterministic xApp-like executor.

\noindent\textbf{Listing A.4: Example LLM-generated YAML policy accepted during the ns-3 campaign.}

\begin{lstlisting}[style=agenticv2x]
policy_id: qwen2_5_7b_real_100s_20veh_seed6_rapp_policy
description: Conservative rApp policy for V2X scheduler weights
scenario_id: real_100s_20veh_seed6
objective_id: safety_background_balance
baseline_weights:
  ToD: 6
  Awareness: 5
  HDMap: 3
  Sensor: 2
bounds:
  ToD:
    - 4
    - 8
  Awareness:
    - 4
    - 8
  HDMap:
    - 1
    - 7
  Sensor:
    - 1
    - 7
safety_constraints:
  min_critical_weight: 5
  max_background_degradation: 0.35
  forbid_zero_weight: true
  critical_services:
    - ToD
    - Awareness
xapp_rules:
  - name: tod_latency_guard
    condition:
      metric: tod_p95_latency_ms
      operator: ">"
      value: 20
    action:
      service: ToD
      type: increase_weight
      value: 1
  - name: awareness_violation_guard
    condition:
      metric: awareness_violation_rate
      operator: ">"
      value: 0.05
    action:
      service: Awareness
      type: increase_weight
      value: 1
  - name: congestion_hdmap_trim
    condition:
      metric: congestion_ratio
      operator: ">"
      value: 0.7
    action:
      service: HDMap
      type: decrease_weight
      value: 1
  - name: congestion_sensor_trim
    condition:
      metric: congestion_ratio
      operator: ">"
      value: 0.8
    action:
      service: Sensor
      type: decrease_weight
      value: 1
fallback_policy: static_expert
runtime_safety:
  rollback_enabled: true
  degradation_windows: 3
  monitored_metrics:
    - critical_violation_rate
    - critical_dc_prr
    - critical_p95_latency_ms
  directional_guard_enabled: true
  max_ul_dl_dc_prr_gap: 0.05
metadata:
  source: llm_rapp
  model_name: qwen2.5:7b
  generation_time: null
\end{lstlisting}

%

\end{document}